\documentclass[epj]{svjour}
\usepackage{graphics}
\usepackage{color}
\usepackage[greek,english]{babel}
\newcommand{\gP}{\greektext P\latintext}

\newcommand{\gD}{\greektext D\latintext}
\usepackage[square,numbers]{natbib}
\begin{document}

\title{Measuring the electric dipole moment of the electron in BaF}

\author{The NL-$e$EDM collaboration: Parul Aggarwal\inst{1}, Hendrick L. Bethlem\inst{2}, Anastasia Borschevsky\inst{1}, Malika Denis\inst{1}, Kevin Esajas\inst{1}, Pi A. B. Haase\inst{1}, Yongliang Hao\inst{1}, Steven Hoekstra\inst{1} \thanks{\emph{corresponding author: s.hoekstra@rug.nl}}, Klaus Jungmann\inst{1}, Thomas B. Meijknecht\inst{1}, Maarten C. Mooij\inst{2}, Rob G. E. Timmermans\inst{1}, Wim Ubachs\inst{2}, Lorenz Willmann\inst{1}, Artem Zapara\inst{1}%
}
\institute{Van Swinderen Institute for Particle Physics and Gravity (VSI), University of Groningen, Groningen, The Netherlands, and Nikhef, National Institute for Subatomic Physics, Science Park 105, 1098 XG Amsterdam, The Netherlands \and  LaserLaB, Department of Physics and Astronomy, Vrije Universiteit, De Boelelaan 1081, 1081 HV Amsterdam, The Netherlands }

\date{\today}

\abstract{
We investigate the merits of a measurement of the permanent electric dipole moment of the electron ($e$EDM) with barium monofluoride molecules, thereby searching for phenomena of CP violation beyond those incorporated in the Standard Model of particle physics. Although the BaF molecule has a smaller enhancement factor in terms of the effective electric field than other molecules used in current studies (YbF, ThO and ThF$^+$), we show that a competitive measurement is possible by combining Stark-deceleration,  laser-cooling and an intense primary cold source of BaF molecules. With the long coherent interaction times obtainable in a cold beam of BaF, a sensitivity of $5\times10^{-30}$ e$\cdot$cm for an $e$EDM is feasible. We describe the rationale, the challenges and the experimental methods envisioned to achieve this target.
} %end of abstract

\authorrunning{Aggarwal \textit{et al.}}
\titlerunning{Measuring the electric dipole moment of the electron in BaF}
\maketitle
\section{Introduction}
\label{intro}
The discovery of the Higgs boson~\cite{Aad2012,Chatrchyan2012} at CERN's Large Hadron Collider (LHC) started a new era in particle physics. The LHC will continue to test the Standard Model (SM) by mapping out the properties of the Higgs boson with increasing precision. Moreover, now that we know that the Higgs boson exists, unanswered questions in particle physics and cosmology come into focus. The Higgs field is responsible for the electroweak symmetry breaking in the very early Universe. The most natural explanation of why this breaking occurs at a scale of 100 GeV predicts new particles with masses in the TeV range, some of which could make up the dark matter in the Universe~\cite{Bertone2005}. Leptons and quarks acquire mass via the Higgs field, such that particles and antiparticles have slightly different weak interactions. This symmetry breaking between matter and antimatter is called CP violation, where C denotes charge conjugation and P parity. CP violation within the SM, however, is insufficient, by orders of magnitude, to explain the matter-antimatter asymmetry in the Universe. This failure of the SM provides a major motivation to search for new sources of CP violation at the TeV scale. This is the mission of, for example, the LHCb experiment~\cite{Aaij2015}.

Experiments that search for permanent electric dipole moments (EDMs) are complementary to, and compete with, searches for CP violation with antiparticles at high-energy colliders~\cite{Pospelov2005}. A nonzero electron EDM ($e$EDM) implies the electron effectively has an aspherical charge distribution along its spin axis, which is forbidden by time-reversal (T) invariance. According to the CPT theorem of quantum field theory, T violation is equivalent to CP violation. In the SM the $e$EDM is zero up to the three-loop level. Its predicted value~\cite{Pospelov2005} is $d_e = {\cal O}(10^{-38}$)e$\cdot$cm, which is experimentally far out of reach. However, extensions of the SM invariably predict much larger values, and thus provide a window of opportunity for experimental searches. The reason is that in such theories new heavy particles are introduced that have interactions with CP-violating quantum-mechanical phases, giving rise to an $e$EDM at the one- or two-loop level, or even already at tree level~\cite{Czarnecki2009}. For instance, supersymmetry (SUSY), a conjectured symmetry between bosons and fermions, predicts an $e$EDM due to interactions with virtual selectrons and photinos, the SUSY partners of the electron and photon, at the one-loop level. If such new particles with TeV-scale masses exist, the $e$EDM will have a measurable value in the ongoing and upcoming experiments.

\begin{figure*}[t]
\centering
\resizebox{0.75\textwidth}{!}{%
  \includegraphics{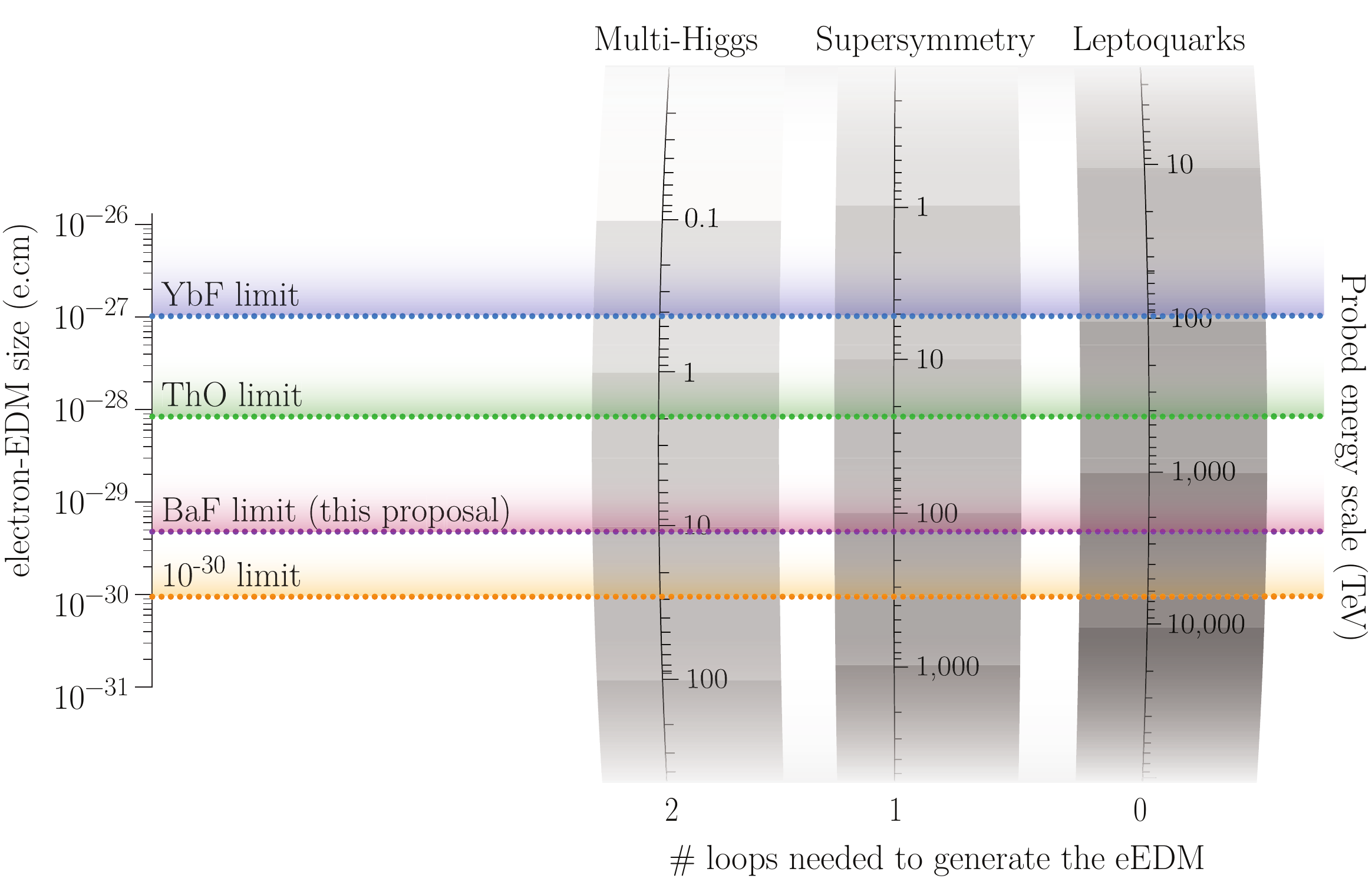}}
\caption{The experimental upper limits on the value of the electron EDM translated into a limit on the probed energy scale through $d_e = e \alpha^n \sin\phi\,(m_e/\Lambda^2)$. Here $\alpha$ is the fine-structure constant, $n$ is the number of loops needed to generate the $e$EDM. $\phi$ is a typical CP-violating phase, which is set to the maximum value of $\pi/2$, and $\Lambda$ is the energy scale that is being probed. The regions above the dotted lines are the energy ranges that are excluded for the indicated classes of models. The models differ in the number of loops required to generate an $e$EDM. The $e$EDM value predicted by the Standard Model, in which the $e$EDM only appears at 4-loop level, is of the order of $10^{-38}$ e$\cdot$cm.}
\label{fig:limits}
\end{figure*}

The relation between the masses of the yet to be discovered particles and the $e$EDM value in different types of models is illustrated in Fig.~\ref{fig:limits}. The blue and green lines show the limits on the $e$EDM set by experiments with molecular beams of YbF~\cite{Hudson2011} and ThO~\cite{Baron2014}, respectively. The purple line shows the limit aimed at in our experiment with BaF. It is obvious from Fig.~\ref{fig:limits} that existing $e$EDM experiments put strong constraints on the validity of the SUSY models and other speculative SM extensions. Improved $e$EDM experiments will probe energies approaching the PeV scale and will thereby contribute to the roadmap of particle physics beyond the LHC. 

In this article, we outline our plans to construct and perform an experiment to search for an $e$EDM with an intense cold beam of barium monofluoride (BaF) molecules. Our program exploits novel techniques to manipulate and control the quantum states of molecules with electric, magnetic and light fields. Such methods have in recent years promoted diatomic molecules to ultrasensitive probes of particle physics \cite{Demille2015, Demille2017}. We show that this experiment will be sensitive to an $e$EDM value of $d_e$ = 5$\times$(10$^{-30})$ e$\cdot$cm, which is an improvement by more than one order of magnitude compared to the current upper limit.

\section{The ${\bf e}$EDM in atoms and molecules}
\label{stateofart}
\textbf{}

Searches for permanent dipole moments of fundamental particles started already before the discovery of P and CP violation, with the investigation of the neutron \cite{Purcell1950}. Ref. \cite{Sandars1967} contains an early proposal to measure the EDM of the proton. In later years, atomic species such as mercury \cite{Griffith2009}, thallium \cite{Regan2002}, and xenon \cite{Rosenberry2001} were employed in dedicated searches for a permanent EDM and the accompanying CP violation, and the neutron also remained at the focus of attention \cite{Baker2006}. For a historical overview of this field of research  we refer to Ref. \cite{Jungmann2013}.

The EDM of composite systems, such as atoms or molecules, can originate in nuclear or electronic contributions (or both), depending on the  electronic structure. In addition, in such systems, large enhancement factors can occur for the particle EDMs due to the strong internal electric field near a heavy nucleus. This enhancement is denoted the effective electric field ($E_{\mathrm{eff}}$). In some atoms these enhancements can reach values of $10^3$-$10^4$, while in molecules they can go up to $10^6$~\cite{Dzuba2012}. Such systems therefore form sensitive testing grounds for probing new sources of CP violation. In paramagnetic systems, CP violation gives rise not only to an $e$EDM, which is usually the focus of the investigations, but also to semileptonic CP-violating interactions between the electrons and the quarks in the atomic nuclei. Theoretical tools are needed to disentangle the different contributions and to relate these to CP-violating observables measured at colliders~\cite{Ginges2004,Engel2013}.

In recent years, experiments with neutral diatomics have set new stringent experimental $e$EDM limits. The Hinds group at Imperial College, London, uses a supersonic beam of YbF molecules \cite{Hudson2011}; YbF is predicted to have $E_{\mathrm{eff}}$ of 23 GV/cm \cite{Abe2014}. The ACME collaboration at Yale/Harvard uses a buffer-gas beam of metastable ThO molecules \cite{Baron2014}, exploiting the large predicted $E_{\mathrm{eff}}$ in the range of 75-84 GV/cm \cite{Meyer2008,Skripnikov2013,Fleig2014,Skripnikov2015,Skripnikov2016,Denis2016}. The latter experiment was sensitive to frequency shifts of $< 6$ mHz or an energy shift of $< 3\times10^{-18}$ eV (3 aeV), which can be translated, knowing the $E_{\mathrm{eff}}$, to a limit on the electron-EDM of $|d_e| < 8.7\times10^{-29}$ e$\cdot$cm. This is currently the most stringent constraint on the electron EDM. Other neutral species have also been considered for electron EDM searches, such as the WC \cite{Lee2013} and the PbO \cite{Eckel2013} molecules.
There is also an endeavor to use trapped molecular ions \cite{Loh2013} and a study based on the $^{180}$Hf$^{19}$F$^+$ ion has led to a constraint of $|d_e| < 1.3\times10^{-28}$ e$\cdot$cm \cite{Cairncross2017}.

We aim to use the neutral BaF molecule for an electron EDM measurement. In heavy paramagnetic diatomic molecules such as BaF the valence electron is exposed to a huge internal electric field, parametrized by $E_{\mathrm{eff}}$, which enhances the effect of the $e$EDM and results in a linear Stark shift. This shift is the experimental signal that we aim to measure \cite{Hinds1997}. The magnitude of $E_{\mathrm{eff}}$ cannot be measured directly and has to be calculated using electronic structure methods. BaF has a somewhat smaller predicted $E_{\mathrm{eff}}$ of 6-8.5 GV/cm compared to other molecules used in such experiments, depending on the computational approach \cite{Kozlov1995,Nayak2006,Meyer2008,Fukuda2016,Gaul2017, Abe2018}. We are currently performing benchmark quality calculations of the $E_{\mathrm{eff}}$ of BaF, which is needed for the interpretation of the measurements and extraction of the limit on the electron EDM. The relativistic coupled cluster approach (CCSD(T)), considered to be one of the most powerful computational methods, is used in these calculations, and we are also investigating the influence of various computational parameters on the value of obtained $E_{\mathrm{eff}}$ \cite{Haase2018}. The modest value of $E_{\mathrm{eff}}$  is however compensated by the fact that BaF is lighter than YbF and ThO, which allows it to be efficiently decelerated in a molecular beam machine of realistic size. Furthermore, unlike ThO, the experiment will be performed in the electronic ground state that has a suitable structure for efficient laser cooling and detection.

\section{Scientific challenges}
\label{challenges}
\subsection{Boosting the sensitivity of an $e$EDM measurement}
\begin{figure}
\resizebox{0.5\textwidth}{!}{%
  \includegraphics{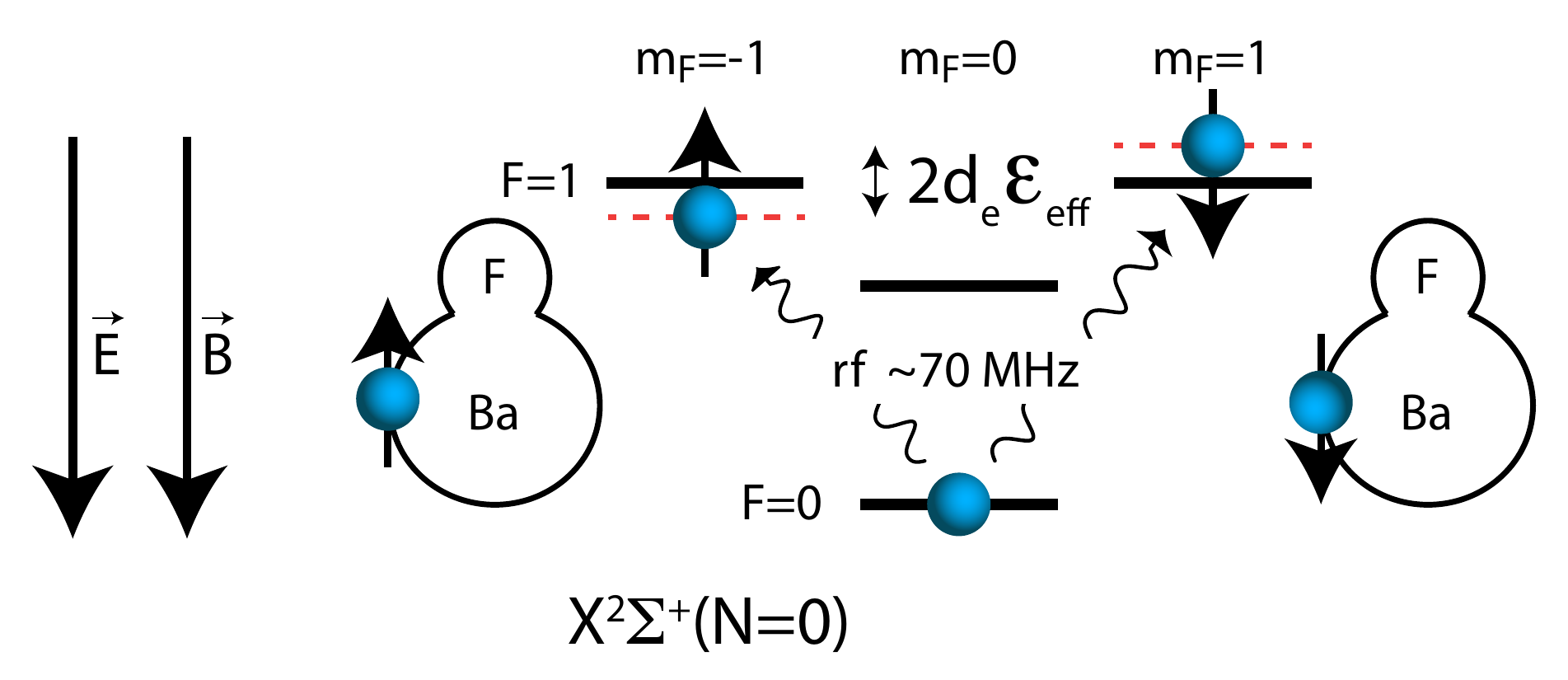}
}
\caption{The ingredients of an $e$EDM measurement in BaF molecules. A superposition of hyperfine substates is created, which builds up a phase difference during the interaction with the electric and magnetic fields. This phase difference, proportional to the product of the $e$EDM magnitude and the effective electric field inside the molecule, is read out when the superposition is projected back onto the $N=0$, $F=0$ state.}
\label{fig:ingredients}
\end{figure}

The experimental strategy to search for a finite EDM of the electron is illustrated in Fig.~\ref{fig:ingredients}. A superposition of hyperfine substates is created, which builds up a phase difference during the interaction with external electric and magnetic fields. The EDM signal is detectable through a difference in the total accumulated phase for the parallel and the anti-parallel orientation of the magnetic and electric fields. The phase difference $\phi_{\mathrm{EDM}}$ due to the EDM is measured by subtracting the average fluorescence count rates for the relative orientations, and is directly proportional to the $e$EDM
\begin{equation}
\phi_{\mathrm{EDM}} = d_e |P| E_{\mathrm{eff}}\tau/\hbar~~~,
\end{equation}
where $P$ is the polarisation factor of the molecule and $\tau$ is the coherent interaction time in the interaction zone. The experimental resolution that can be reached depends on the fringe contrast, i.e. on the statistics, on the stability and homogeneity of the applied electric and magnetic fields, as well as on the velocity and flux stability of the molecular beam.

The statistical uncertainty $\sigma_d$ of a molecular $e$EDM experiment is determined by the interplay of four crucial parameters: the rate of detected molecules $\dot N = dN/dt$, the coherent interaction time of the molecules with the electric field $\tau$, the measurement time $T$, and $|P| E_{\mathrm{eff}}$, and is given by
\begin{equation}
\label{statistics}
\sigma_d = \frac{\hbar}{e}\frac{1}{2 |P| E_{\mathrm{eff}} \tau \sqrt{\dot N T}}~~~.
\end{equation}
It is attractive to use a long coherent interaction time $\tau$, as the sensitivity improves linearly with this parameter. Until recently, increasing the interaction time was accompanied by a significant decrease in the counting rate $\dot N$. However, recent advances in decelerating molecular beams \cite{Osterwalder2010, Bulleid2012, Quintero-perez2014,Vandenberg2014} combined with spectacular progress in molecular laser cooling \cite{Shuman2010} and the demonstration of intense cryogenic molecular beam sources \cite{Patterson2007} have opened a route to circumvent this limitation and make long interaction times possible. A similar approach has recently been suggested for the YbF molecule \cite{Lim2017}.

\begin{figure*}
\resizebox{1\textwidth}{!}{%
  \includegraphics{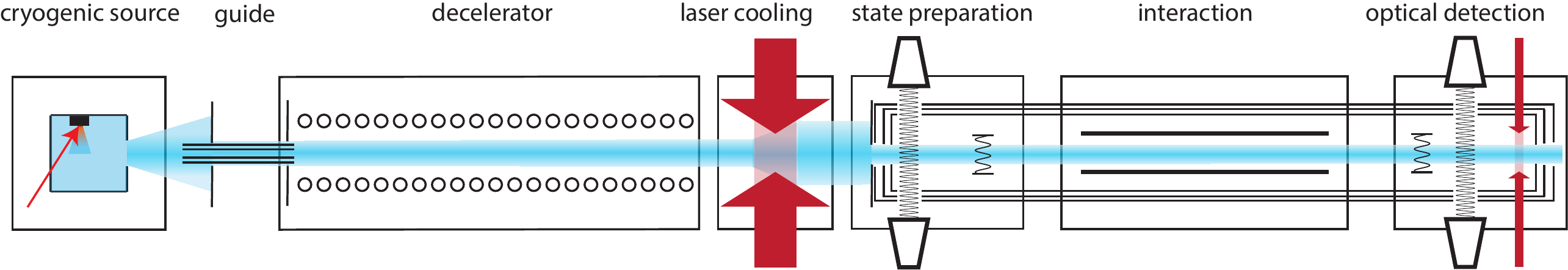}
          }
\caption{Schematic overview of the proposed experimental approach. The molecular beam of BaF molecules travels from the left to the right. A beam with a velocity of $\sim180$ m/s is created in the cryogenic source, and subsequently decelerated to $\sim30$ m/s in the Stark decelerator. A lasercooling section reduces the transverse velocity spread of the beam, preventing it from spreading out during the $\sim15$ ms it takes the molecules to travel through the 0.5 m long interaction zone. In the magnetically shielded interaction zone the $e$EDM is probed using Ramsey interferometry in an electric field. Readout of the signal is done by fluorescence detection.}

\label{fig:overview}
\end{figure*}

Our experimental strategy is summarized in Fig.~\ref{fig:overview}. The use of advanced slowing and cooling techniques creates a slow and well-collimated molecular beam, which leads to a coherent interaction time one order of magnitude longer than that of competing experiments, without sacrificing the counting rate. Slow BaF molecules can be detected efficiently by exploiting the closed cycling transition, which further boosts the sensitivity.

We made a conservative estimate of the statistical sensitivity that we can obtain, resulting in $5\times10^{-30}$ e$\cdot$cm. This is based on the detection of $7\times10^5$ molecules/shot, with the experiment running at 10 Hz for a period of 24 hours. All essential numbers leading to this estimate are summarized in Table~\ref{tab:1}. In section~\ref{sec:setup} the four main components of the proposed experiment are described in detail, and more information is given to support the numbers in the table.

\begin{table*}[t]
\caption{The estimate of the number of molecules that can be detected per repetition of the experiment. We aim to run the experiment at 10 Hz.}
\label{tab:1}       
\begin{tabular*}{\textwidth}{l @{\extracolsep{\fill}} p{25 pt}p{180 pt}p{180 pt}}
\hline\noalign{\smallskip}
Item & Number & Units & Resulting \# mol./shot  \\
\noalign{\smallskip}\hline\noalign{\smallskip}
Source & $10^{13}$ & Molecules/shot &  \\
 & 0.005 & Extraction efficiency from buffer gas cell &  \\
 & 0.24 & Fraction in $v=0$, $N=2$ & $5\times10^{10}$ from source; $4\times10^9$  in desired state,  \\
 & 0.3 & Fraction in low-field seeking states & v$_{\mathrm{long}}$=($180\pm50$) m/s, v$_{\mathrm{trans}}$=$\pm30$ m/s. \\
 \hline\noalign{\smallskip}
Decelerator & 0.002 & Fraction in velocity acceptance &  \\
 & 0.3 & Fraction in spatial acceptance &  \\
 & 0.7 & Efficiency of deceleration relative to guiding & $2\times10^6$, v$_{\mathrm{long}}$=($30\pm6$) m/s, v$_{\mathrm{trans}}$=$\pm5$ m/s. \\
 \hline\noalign{\smallskip}
Laser cooling & 0.8 & Laser cooling efficiency &  \\
 & 0.7 & State transfer efficiency & $9\times10^5$,  v$_{\mathrm{long}}$=($30\pm6$) m/s, v$_{\mathrm{trans}}$=$\pm0.2$ m/s. \\
 \hline\noalign{\smallskip}
Interaction zone & 0.8 & Transmission and state transfer efficiency &  \\
 & 1.0 & Detection efficiency & $7\times10^5$ \\
\noalign{\smallskip}\hline
\end{tabular*}
\end{table*}

\subsection{Systematic effects}
The achievable limit on the EDM of the electron depends not only on the statistical uncertainty (see eq. \ref{statistics}) but also on the control and understanding of experimental procedures, i.e, on systematic uncertainties. The measurement principle and the interaction zone are similar to that of the ongoing molecular EDM experiments such as ACME at Harvard \cite{Baron2017} and the YbF experiment at Imperial College \cite{Kara2012}. The systematic effects are thus comparable; however, they need to be evaluated explicitly in the context of the current BaF experiment. We expect to control the combined systematic effects at a level corresponding to an $e$EDM sensitivity of $10^{-30}$ e$\cdot$cm.

In order to suppress systematics all experimental parameters, such as magnetic and electric fields as well as rf and light polarizations, will be switched (reversed)~\cite{Hudson2014}. Such reversals can in practice be achieved with finite accuracy. For example, reversing the electric field may result in a slightly different field strength, which can cause an asymmetry in the experimental conditions. As imperfections in these fields can mimic EDM signals, their mapping and in situ monitoring will be implemented by electro-optical field sensors. Systematic effects that cannot be canceled by parameter reversal need to be carefully addressed individually. These effects will be studied making use of the well controlled molecular beam and, where possible, taken into account by their deliberate exaggeration.

A major distinction of the current approach compared to the previous $e$EDM experiments is the long interaction time, which results in increased demands on the magnetic field control. The low value of the magnetic holding field ($\sim 600$ pT) requires careful shielding of both static and dynamic stray magnetic fields. For this the background magnetic field will be compensated (to about 5\%) by 3 sets of large electric current driven field coils around the interaction zone setup. The interaction zone itself will be embedded in a cylindrical multilayer $\mu$-metal shield for primarily static field compensation. A coaxial, few mm thick Al shield will be used for AC environmental field suppression, with particular emphasis on the AC field frequencies with periods comparable to the transit time of the molecules.

\section{Experimental set-up}\label{sec:setup}

\begin{figure*}
\resizebox{1\textwidth}{!}{%
  \includegraphics{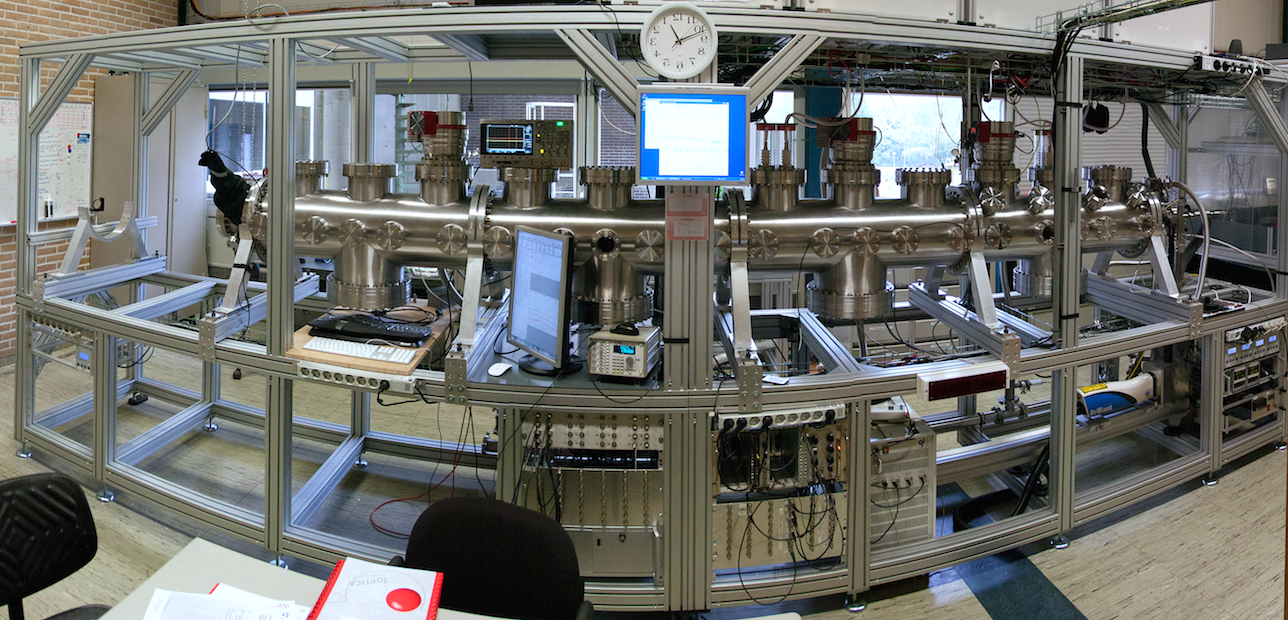}
           }
\caption{Panoramic view of the Stark decelerator, operational at VSI in Groningen. This device, with a length of 4.5 meter, was built with the purpose of studying fundamental symmetry violations with slows beams of heavy diatomic molecules.}
\label{fig:panoramic}       % Give a unique label
\end{figure*}

\subsection{Cryogenic source}
Until now all Stark deceleration experiments (including the experiments with SrF at the Van Swinderen Institute (VSI) at the University of Groningen \cite{Vandenberg2014, Mathavan2016}) have used pulsed supersonic beams as the source. However, over the last decade several groups have demonstrated a promising new type of molecular-beam source using the cryogenic buffer gas method \cite{Patterson2007,Hutzler2011,Bulleid2013, Maxwell2005, vanBuuren2009}. With this method, cold beams of refractory species are generated by ablating a solid precursor inside a cold cell. This cold cell, with typical dimensions of a few centimeters, is filled with helium or neon atoms cooled to $2 - 20$ K by a closed-cycle cryocooler. The buffer gas in the cell is kept at a specifically tuned atom number density, which is low enough to prevent simple three-body collision cluster formation yet high enough to provide enough collisions for thermalization before the molecules touch the walls of the cold cell. A beam of cold molecules can be formed when the buffer gas and target molecules escape the cell through a few-millimeter sized opening into a high vacuum region. When the helium or neon flow through the cell is small, an effusive beam is formed with a relatively low forward velocity, which scales as the square root of the temperature over the mass of the molecule. At higher flow rates the beam velocity increases, eventually becoming equal to the square root of the temperature over the mass of the atom used as buffer gas, while the extraction efficiency increases as well and the beam becomes more collimated. Beams of slow ($<200$ m/s) heavy molecules, such as SrF, BaF, YbF and ThO, with a brightness of above $10^{11}$ molecules per steradian per pulse have been reported \cite{Skoff2011,Hutzler2012}.

We will use a cryogenic buffer gas source with a design similar to that demonstrated by Hutzler et al. \cite{Hutzler2012} and Bulleid et al. \cite{Bulleid2013}, to create pulses of BaF molecules with a forward velocity of around 180 m/s. At these velocities, the cryogenic source will generate an intense beam of forward-directed molecules that can be effectively captured by our traveling wave decelerator. As a conservative estimate, $10^{13}$ molecules are created per ablation pulse, of which a fraction of $5\times10^{-3}$ is extracted into the molecular beam. We therefore expect to be able to create a beam with $5\times10^{10}$ molecules in a pulse of $5-10$ ms with a rotational temperature of 2 K, a longitudinal velocity spread of 100 m/s, and a transverse velocity spread of 60 m/s. This intensity is similar to that demonstrated for BaF by Zhou \cite{Zhou2015}, and about a factor 10 smaller than demonstrated for ThO \cite{Hutzler2011}.

In order to prevent the buffer gas from entering the decelerator beamline, we will add a series of quadrupole lenses of design similar to those used in the molecular fountain and molecular synchrotron experiments at the Vrije Universiteit, Amsterdam ~\cite{Cheng2016, Zieger2010}, which will guide the BaF molecules from the exit of the cryogenic source to the entrance of the decelerator over a distance of 0.5 to 1 meter. By adding some length, the coupling to the longitudinal phase-space matching of the decelerator can be improved~\cite{Fabrikant2014}. The guide has a much larger transverse acceptance than the decelerator, so molecules that are lost transversally in the guide would not have been decelerated anyway. The last section of the guide will be used to match the transverse phase-space distribution of the beam to the acceptance of the decelerator \cite{Meerakker2012}.

\begin{figure*}
\resizebox{0.47\textwidth}{!}{%
  \includegraphics{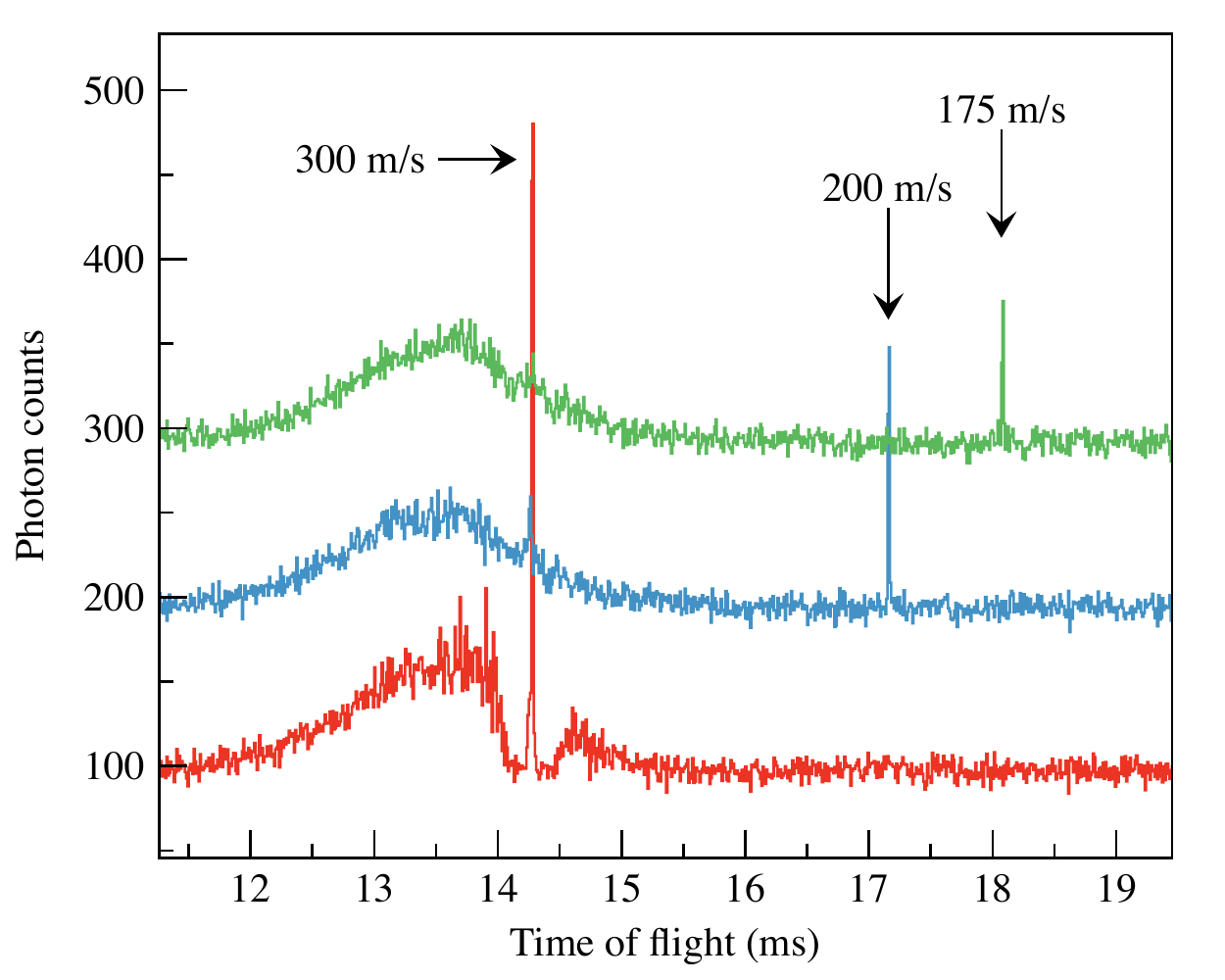}
          }
\resizebox{0.5\textwidth}{!}{%
  \includegraphics{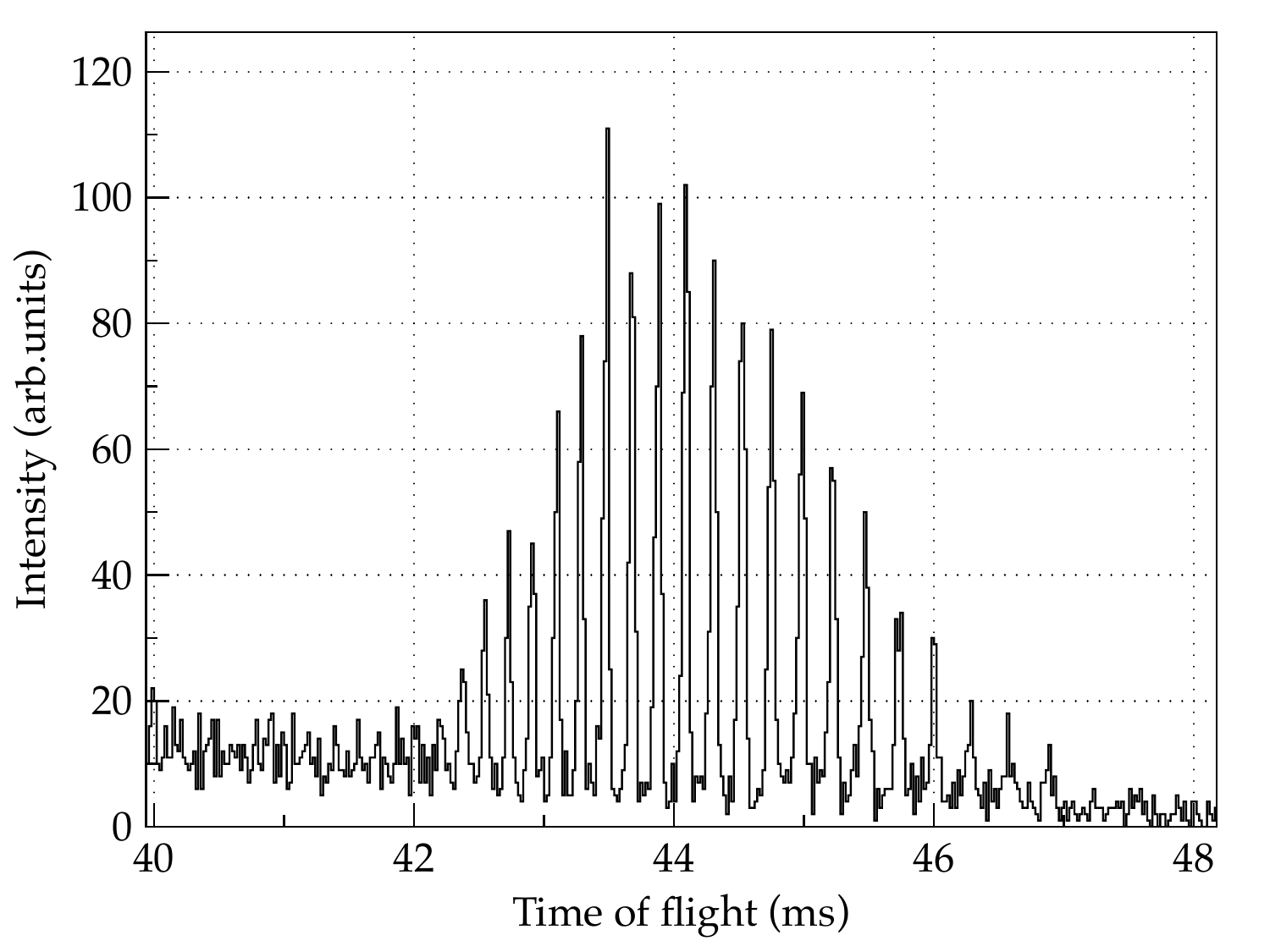}
          }
\caption{(a) Time of flight profile showing the arrival time of SrF molecules captured from a supersonic expansion when a waveform is applied to guide (red) or decelerate (blue and green) the molecules \cite{Vandenberg2014,Mathavan2016}. These measurements are performed with a 4 meter long decelerator. The central peak in these plots corresponds to a packet of molecules kept together by the electric fields throughout the deceleration process. (b) A numerical simulation of the time of flight profile illustrates the deceleration of BaF molecules from a cryogenic source, from 180 to 30 m/s, in a 4.5 meter long decelerator. Molecules from the long pulse are captured into neighboring traps of the traveling-wave decelerator, resulting in multiple peaks in the time of flight profile. All decelerated molecules can be used for the EDM measurement because they will all propagate, after deceleration, through the interaction zone at ($30\pm6$) m/s.}
\label{fig:TOF}       % Give a unique label
\end{figure*}

\subsection{Stark deceleration} 
To reduce the forward velocity of the molecular beam we will use a traveling-wave decelerator. This decelerator (depicted in Fig.~\ref{fig:panoramic}) for heavy diatomic molecules was recently built and taken into operation at VSI. In recent years we have performed a number of experiments to prepare samples of molecules for precision measurements using traveling-wave decelerators ~\cite{VandenBerg2012,Quintero-perez2013,Jansen2013,Vandenberg2014, Quintero-perez2014, Mathavan2016}.

The traveling-wave decelerator consists of many ring-shaped electrodes that effectively form a 4 mm diameter tube through which the molecular beam travels. Recent results, shown in Fig.~\ref{fig:TOF}(a), illustrate the efficient deceleration of SrF molecules from a supersonic source with this device. From these results, combined with numerical trajectory simulations, we can deduce the deceleration efficiency for BaF molecules, illustrated in Fig.~\ref{fig:TOF}(b). Due to its length, the VSI decelerator can operate at modest deceleration strengths, which results in the deceleration of the heavy BaF molecules with relatively high efficiency. In this process the molecules are kept together in closely spaced traps of 3 by 6 mm at all times. As the cryogenic beam emits a rather long pulse, multiple co-moving traps inside the decelerator will be filled. The molecules in these traps will all be decelerated to the same final speed and will all contribute to the $e$EDM measurement.

\begin{figure}
\resizebox{0.5\textwidth}{!}{%
  \includegraphics{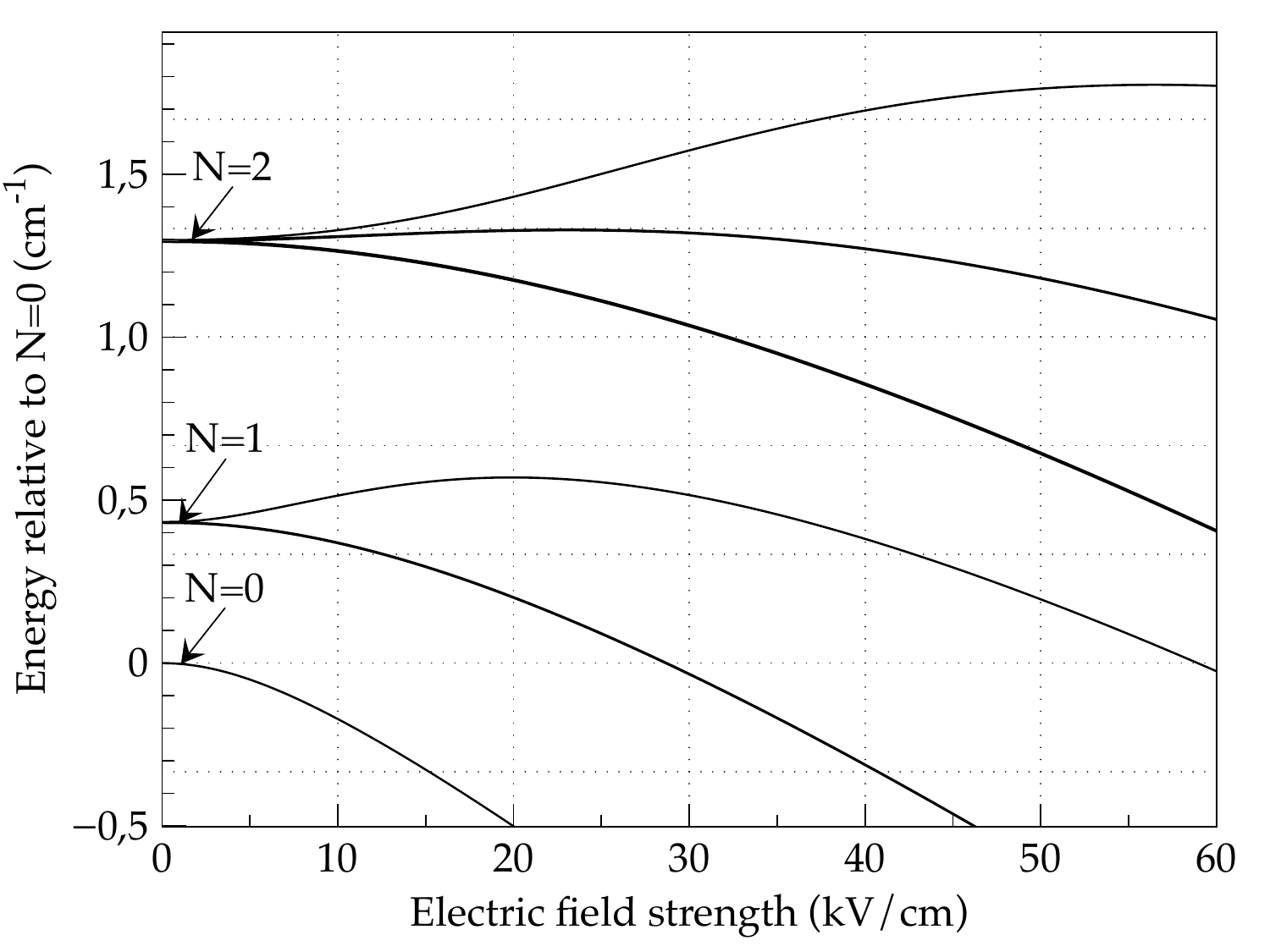}
          }
\caption{The lowest three rotational levels within the electronic ground state of the BaF molecule, as a function of electric field strength. The $N=0$ level is used for the $e$EDM measurement; deceleration is most efficient in the $N=2$ level, while lasercooling and detection is best performed in the $N=1$ level.}
\label{fig:baf_stark}       % Give a unique label
\end{figure}

The number of molecules accepted by the decelerator is determined by the required deceleration strength, the applied voltage, and the Stark shift of the molecular state of interest. The Stark shift of the lowest three rotational levels in the electronic ground state of BaF is shown in Fig.~\ref{fig:baf_stark}. The low-field seeking states of BaF have a turning point in their Stark shift that limits the accepted transverse and longitudinal velocities, which makes deceleration of molecules in the first excited rotational state ($N=1$) optimal at 5 kV, and deceleration of molecules in the $N=2$ state optimal at 10 kV. At 5 kV, BaF molecules in the $N=1$ state are accepted if they enter the decelerator with a longitudinal velocity within $\pm5$ m/s of the set velocity and a transverse velocity up to $\pm3$ m/s. At an increased voltage amplitude of 10 kV, BaF molecules in the $N=2$ state are accepted if they enter the decelerator with a longitudinal velocity within $\pm8.5$ m/s of the set velocity and a transverse velocity up to ±5 m/s. In order to decelerate a beam of BaF in the $N=2$ state from 180 m/s to 30 m/s in 4.5 meters, the traveling well will be decelerated at 3.5 km/s$^2$, which reduces the longitudinal acceptance of the decelerator to 70\% compared to the situation where the well is traveling at a constant speed. Assuming a cryogenic beam of BaF with a density and velocity spread as discussed in the previous section, we find that a fraction of $4\times10^{-4}$ of the cryogenic beam can be decelerated. Consequently, the decelerated beam exiting the traveling-wave decelerator will contain $2\times10^6$ molecules per pulse, with a longitudinal velocity spread of $\pm6$ m/s and a transverse velocity spread of $\pm5$ m/s.

\subsection{Laser cooling} 
The slow molecular beam exiting the decelerator diverges due to the transverse velocity spread. To allow the molecular beam to pass through the 50 cm long EDM measurement zone, laser cooling will be employed to reduce the transverse velocity component. Since laser cooling requires the molecules to be in the $N=1$ level, the molecules will be transferred from the low-field seeking $F=2$ and $F=3$ states of the $N=2$ level to the $F=1$ and $F=2$ states of the $N=1$ using microwave radiation at 25.2 GHz, at the beginning of the laser cooling section.

\begin{figure}
\resizebox{0.5\textwidth}{!}{%
  \includegraphics{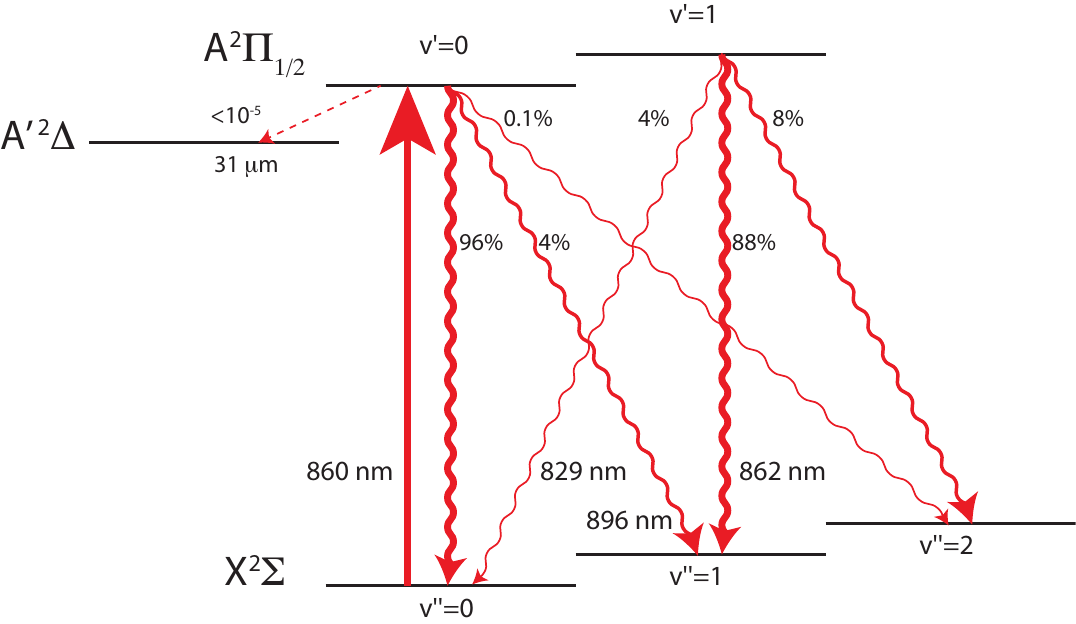}
          }
\caption{The lowest vibrational levels within the electronic ground state and electronically excited A$^2$\gP$_{1/2}$ state of the BaF molecule. The wavelengths of the main transitions are taken from experiment \cite{Bernard1992}; the branching ratios from the excited state are from calculations performed in our group \cite{Hao2018}. Given the low number of photons required for the cooling, two additional lasers repumping molecules from the $v^{\prime\prime}=1$ and $v^{\prime\prime}=2$ are sufficient.}
\label{fig:levels}       % Give a unique label
\end{figure}

The powerful technique of laser cooling has only recently been extended from atoms to molecules \cite{Shuman2010,Hummon2013,Zhelyazkova2014, Barry2014, Kozyryev2017}, where the leaky optical cooling cycle leads to challenges comparable to the laser cooling of atoms like Ba \cite{De2009}. Some molecules though, like BaF, have however a number of properties that make them well-suited for laser cooling, that is i) a low probability for vibrational excitations upon electronic excitation, ii) a short excited-state lifetime \cite{Berg1993,Berg1998}, which is important for a fast cooling cycle, and iii) convenient transition wavelengths~\cite{Effantin1987,Effantin1990,Bernard1990,Bernard1992}, where diode lasers with sufficient intensity are available. Recently, a number of investigations have highlighted the possibility of lasercooling of BaF molecules~\cite{Chen2016,Chen2017}. We have carried out relativistic coupled cluster calculations of the branching ratios and relativistic multireference configuration interaction (MRCI) calculations of the transition dipole moments of the relevant transitions~\cite{Hao2018}. The obtained Franck-Condon factors are indicated along with the experimentally known energy level structure of BaF in Fig.~\ref{fig:levels}. One main cooling transition and repumping from the vibrational states $v^{\prime\prime}=1$ and $v^{\prime\prime}=2$ is sufficient for the transverse cooling of the BaF beam. The branching to the Delta state A$^{\prime 2}$\gD~is strongly suppressed due to the small energy separation~\cite{Barrow1988,Kang2016}. Frequency sidebands will be created on both the main cooling transition and the repump transition to match the hyperfine structure in the ground state~\cite{Ernst1986}.

The efficient transverse laser cooling is possible because of the advantageous properties of the slowed BaF beam at the exit of the decelerator. A single photon at 860 nm results in a recoil of BaF of about 3 mm/s. In 2D optical molasses, scattering of 2000 photons suffices to dissipate the 5 m/s transverse velocity at the exit of the decelerator, taking less than 2 ms at saturation intensity. During the transverse cooling the molecules travel $\sim6$ cm and the beam diameter increases to $\sim1$ cm. Saturation intensity in the area of the transverse cooling requires on the order of 100 mW of laser power. Conveniently, the required laser wavelengths and intensities are all available as diode lasers with tapered amplifiers. In order to cover the hyperfine structure splitting of about 25 MHz, sidebands to the laser frequency will be generated by electro-optical and acousto-optical modulation or by frequency offset locking of individual diode lasers. The absolute wavelength of the light will be stabilized against the optical frequency comb at the VSI for stable long-term operation. From simulations using the calculated branching ratios, we find that less than 20\% of the molecules are lost to states other than the cooling state. 

After being laser-cooled, the molecules will be optically pumped into a single hyperfine state of the $N=1$ level, and subsequently transferred to the $F=0$ hyperfine level of the $N=0$ rotational ground state by resonant microwave radiation at 12.6 GHz. The combined efficiency of the microwave transfers and optical pumping to the EDM state is estimated to be better than 0.7. Hence we expect the molecular beam emerging from the laser cooling section to have $9 \cdot10^5$ molecules/shot, with the transverse velocity reduced to $\pm 0.2$ m/s.

\subsection{Interaction zone and measurement}
The interaction zone with a length of 0.5 m will be magnetically shielded using multiple cylindrical layers of $\mu$-metal \cite{Paperno2000} with end-caps inside a set of magnetic coils for environmental static field compensation. The external static and dynamic magnetic fields $B_\mathrm{ext}$ will be cancelled to $|B_{\mathrm{ext}}| < 50$ pT. The magnetic field will be monitored with sensitivity at the several 10 fT level \cite{Budker2007} with fluxgate (and if needed SERF) magnetometers at fixed locations outside and inside the shielded volume, based on the results of the magnetic field survey at the experiment site. Of particular importance are dynamic field changes at frequencies originating from laboratory technological equipment (such as, e.g., 50 Hz), the period of which is comparable to the travel time of the molecular beam through the interaction volume. During the measurements we will use a set of additional coils external to the apparatus in the interaction zone as a feedback system for dynamic field compensation to provide sufficient control of environment induced magnetic field changes on fast (sub-second) timescales. In order to avoid phase locking of the data acquisition to the frequency of potential perturbations we will adjust the repetition rate for the measurements to be asynchronous with technical frequencies. These measures also prepare the interaction zone for future experiments with slower beams.

A homogeneous vertical magnetic holding field of $600~{\rm pT}$ corresponds to a spin precession angle of $\pi/4$ when the molecules pass through the interaction zone inside the passive magnetic shield similar to that employed in the $^{129}$Xe EDM experiment ~\cite{Allmendinger2017}. A static field homogeneity of $< 10^{-3}$ can be reached with an additional coil system inside the $\mu$-metal shield for higher magnetic multipole compensation. The magnetic field polarity will be regularly reversed throughout the measurements in order to control and check for systematics.

Two parallel plane electrodes, made from a thin low conductance material deposited on glass, provide an electric field of order 10 kV/cm parallel to the magnetic holding field. For control of systematics the electric field strength will be monitored by a recently developed ultra-low frequency electro-optic sensor~\cite{Grasdijk2018}. The molecular state preparation and state detection regions are located near both ends of and inside the electric field volume. This yields quantitative state identification in BaF via fluorescence detection. For the EDM measurements we will repeatedly reverse the electric field on timescales that are long compared to the coherence time. This procedure will include settling times for field stabilization.

After the molecules pass the electric field region the molecular superposition state is projected back onto the first excited rotational level, from where $\sim 10^3$ photons can be scattered per molecule. These photons are imaged with  1\% efficiency onto a position-sensitive low-noise single-photon detector. A cooled EMCCD camera, with a quantum efficiency of 50\% at 860 nm, is capable of quantitatively detecting, without pileup, the high instantaneous photon rates of above $10^7$ photons per shot, which we expect at the detector. Thus, 5 photons on average will be detected for each molecule, which results in a detection efficiency of close to unity.

\section{Future perspectives}
The proposed use of a slow, intense, and cold molecular beam is, with the current level of technology, a promising approach to measure the $e$EDM and will allow for a measurement that is more than one order of magnitude below the current limit. The great sensitivity of the proposed scheme (mainly) stems from the interaction time that is over ten times longer than used in ongoing experiments. We believe that in future experiments the interaction times can be increased even further. The interaction time in a horizontal beam machine is limited to about 15 ms, as with longer times the beam will fall under gravity and will miss the detection zone. An elegant way to circumvent this is by using a vertical beam machine in fountain geometry. We have recently demonstrated the first-ever fountain for molecules, which enables the study of ammonia molecules in free fall for up to 266 milliseconds~\cite{Cheng2016}. A crucial part of this machine is the travelling-wave decelerator that enables us to manipulate the motion of polar molecules virtually without losses. Using a combination of quadrupole lenses and bunching elements, the slow ammonia beam exiting the decelerator is shaped such that it has a large position spread and a small velocity spread when the molecules are in free fall, while being strongly focused at the detection region. Using a similar lens system to focus the laser cooled BaF beam, it should be possible to create a (quasi-cw) fountain offering interaction times of up to a second without significant loss in flux. Ultimately, the most sensitive experiment could be based on optically trapped molecules. However, significant hurdles still have to be overcome to realize this: a means to accumulate a sufficiently large number of molecules has to be demonstrated as well as cooling to significantly lower temperatures than can currently be obtained.

\bibliography{BaF_eEDM}

\begin{thebibliography}{10}

\bibitem{Aad2012}
G.~Aad et~al. (ATLAS~Collaboration).
\newblock Observation of a new particle in the search for the standard model
  higgs boson with the atlas detector at the {LHC}.
\newblock {\em Phys. Lett. B}, 716:1, 2012.

\bibitem{Chatrchyan2012}
S.~Chatrchyan et~al. (CMS~Collaboration).
\newblock Observation of a new boson at a mass of 125 {GeV} with the {CMS}
  experiment at the {LHC}.
\newblock {\em Phys. Lett. B}, 716:30, 2012.

\bibitem{Bertone2005}
G.~Bertone, D.~Hooper, and J.~Silk.
\newblock Particle dark matter: evidence, candidates and constraints.
\newblock {\em Phys. Rep.}, 405:279, 2005.

\bibitem{Aaij2015}
R.~Aaij et~al.
\newblock {{LHCb} detector performance}.
\newblock {\em Int. J. Mod. Phys. A}, 30:1530022, 2015.

\bibitem{Pospelov2005}
M.~Pospelov and A.~Ritz.
\newblock Electric dipole moments as probes of new physics.
\newblock {\em Annals of Physics}, 318(1):119 -- 169, 2005.

\bibitem{Czarnecki2009}
A.~Czarnecki and W.~J. Marciano.
\newblock Electromagnetic dipole moments and new physics.
\newblock {\em Adv. Ser. Direct. High Energy Phys.}, 20:11, 2009.

\bibitem{Hudson2011}
J.~J. Hudson, D.~M. Kara, I.~J. Smallman, B.~E. Sauer, M.~R. Tarbutt, and E.~A.
  Hinds.
\newblock Improved measurement of the shape of the electron.
\newblock {\em Nature}, 473:493--496, 2011.

\bibitem{Baron2014}
J.~Baron, W.~C. Campbell, D.~DeMille, J.~M. Doyle, G.~Gabrielse, Y.~V.
  Gurevich, P.~W. Hess, N.~R. Hutzler, E.~Kirilov, I.~Kozyryev, B.~R.
  O{\textquoteright}Leary, C.~D. Panda, M.~F. Parsons, E.~S. Petrik, B.~Spaun,
  A.~C. Vutha, and A.~D. West.
\newblock Order of magnitude smaller limit on the electric dipole moment of the
  electron.
\newblock {\em Science}, 343:269--272, 2014.

\bibitem{Demille2015}
D.~DeMille.
\newblock Diatomic molecules, a window onto fundamental physics.
\newblock {\em Phys. Today}, 68(12):34--40, 2015.

\bibitem{Demille2017}
D.~DeMille, J.~M. Doyle, and A.~O. Sushkov.
\newblock {Probing the frontiers of particle physics with tabletop-scale
  experiments}.
\newblock {\em Science}, 357(6):990--994, 2017.

\bibitem{Purcell1950}
E.~M. Purcell and N.~F. Ramsey.
\newblock On the possibility of electric dipole moments for elementary
  particles and nuclei.
\newblock {\em Phys. Rev.}, 78:807--807, 1950.

\bibitem{Sandars1967}
P.~G.~H. Sandars.
\newblock Measurability of the proton electric dipole moment.
\newblock {\em Phys. Rev. Lett.}, 19:1396--1398, 1967.

\bibitem{Griffith2009}
W.~C. Griffith, M.~D. Swallows, T.~H. Loftus, M.~V. Romalis, B.~R. Heckel, and
  E.~N. Fortson.
\newblock Improved limit on the permanent electric dipole moment of
  $^{199}\mathrm{Hg}$.
\newblock {\em Phys. Rev. Lett.}, 102:101601, 2009.

\bibitem{Regan2002}
B.~C. Regan, E.~D. Commins, C.~J. Schmidt, and D.~DeMille.
\newblock New limit on the electron electric dipole moment.
\newblock {\em Phys. Rev. Lett.}, 88:071805, 2002.

\bibitem{Rosenberry2001}
M.~A. Rosenberry and T.~E. Chupp.
\newblock Atomic electric dipole moment measurement using spin exchange pumped
  masers of ${}^{129}\mathrm{Xe}$ and ${}^{3}\mathrm{He}$.
\newblock {\em Phys. Rev. Lett.}, 86:22--25, 2001.

\bibitem{Baker2006}
C.~A. Baker, D.~D. Doyle, P.~Geltenbort, K.~Green, M.~G.~D. van~der Grinten,
  P.~G. Harris, P.~Iaydjiev, S.~N. Ivanov, D.~J.~R. May, J.~M. Pendlebury,
  J.~D. Richardson, D.~Shiers, and K.~F. Smith.
\newblock Improved experimental limit on the electric dipole moment of the
  neutron.
\newblock {\em Phys. Rev. Lett.}, 97:131801, 2006.

\bibitem{Jungmann2013}
K.~Jungmann.
\newblock Searching for electric dipole moments.
\newblock {\em Ann. d. Physik}, 525(8-9):550--564, 2013.

\bibitem{Dzuba2012}
V.~A. Dzuba and V.~V. Flambaum.
\newblock Parity violation and electric dipole moments in atoms and molecules.
\newblock {\em Int. J. Mod. Phys, E}, 21(11):1230010, 2012.

\bibitem{Ginges2004}
J.~S.~M. Ginges and V.~V. Flambaum.
\newblock Violations of fundamental symmetries in atoms and tests of
  unification theories of elementary particles.
\newblock {\em Physics Reports}, 397(2):63--154, 2004.

\bibitem{Engel2013}
J.~Engel, M.~J. Ramsey-Musolf, and U.~van Kolck.
\newblock Electric dipole moments of nucleons, nuclei, and atoms: The standard
  model and beyond.
\newblock {\em Progress in Particle and Nuclear Physics}, 71(Supplement C):21
  -- 74, 2013.

\bibitem{Abe2014}
M.~Abe, G.~Gopakumar, M.~Hada, B.~P. Das, H.~Tatewaki, and D.~Mukherjee.
\newblock Application of relativistic coupled-cluster theory to the effective
  electric field in {YbF}.
\newblock {\em Phys. Rev. A}, 90:022501, 2014.

\bibitem{Meyer2008}
E.~R. Meyer and J.~L. Bohn.
\newblock Prospects for an electron electric-dipole moment search in metastable
  {ThO} and {ThF}$^+$.
\newblock {\em Phys. Rev. A}, 78:010502, 2008.

\bibitem{Skripnikov2013}
L.~V. Skripnikov, A.~N. Petrov, and A.~V. Titov.
\newblock Communication: Theoretical study of {ThO} for the electron electric
  dipole moment search.
\newblock {\em J. Chem. Phys.}, 139(22):221103, 2013.

\bibitem{Fleig2014}
T.~Fleig and M.~K. Nayak.
\newblock Electron electric dipole moment and hyperfine interaction constants
  for {ThO}.
\newblock {\em J. Mol. Spectrosc.}, 300:16 -- 21, 2014.

\bibitem{Skripnikov2015}
L.~V. Skripnikov and A.~V. Titov.
\newblock Theoretical study of thorium monoxide for the electron electric
  dipole moment search: Electronic properties of {H$^3\Delta_1$} in {ThO}.
\newblock {\em J. Chem. Phys.}, 142(2):024301, 2015.

\bibitem{Skripnikov2016}
L.~V. Skripnikov.
\newblock Combined 4-component and relativistic pseudopotential study of {ThO}
  for the electron electric dipole moment search.
\newblock {\em J. Chem. Phys.}, 145(21):214301, 2016.

\bibitem{Denis2016}
M.~Denis and T.~Fleig.
\newblock In search of discrete symmetry violations beyond the standard model:
  Thorium monoxide reloaded.
\newblock {\em J. Chem. Phys.}, 145(21):214307, 2016.

\bibitem{Lee2013}
J.~Lee, J.~Chen, L.~V. Skripnikov, A.~N. Petrov, A.~V. Titov, N.~S. Mosyagin,
  and A.~E. Leanhardt.
\newblock Optical spectroscopy of tungsten carbide for uncertainty analysis in
  electron electric-dipole-moment search.
\newblock {\em Phys. Rev. A}, 87:022516, 2013.

\bibitem{Eckel2013}
S.~Eckel, P.~Hamilton, E.~Kirilov, H.~W. Smith, and D.~DeMille.
\newblock Search for the electron electric dipole moment using
  $\ensuremath{\Omega}$-doublet levels in {PbO}.
\newblock {\em Phys. Rev. A}, 87:052130, 2013.

\bibitem{Loh2013}
H.~Loh, K.~C. Cossel, M.~C. Grau, K.-K. Ni, E.~R. Meyer, J.~L. Bohn, J.~Ye, and
  E.~A. Cornell.
\newblock Precision spectroscopy of polarized molecules in an ion trap.
\newblock {\em Science}, 342(6163):1220--1222, 2013.

\bibitem{Cairncross2017}
W.~B. Cairncross, D.~N. Gresh, M.~Grau, K.~C. Cossel, T.~S. Roussy, Y.~Ni,
  Y.~Zhou, J.~Ye, and E.~A. Cornell.
\newblock Precision measurement of the electron's electric dipole moment using
  trapped molecular ions.
\newblock {\em Phys. Rev. Lett.}, 119:153001, 2017.

\bibitem{Hinds1997}
E.~A. Hinds.
\newblock Testing time reversal symmetry using molecules.
\newblock {\em Physica Scripta}, T70:34, 1997.

\bibitem{Kozlov1995}
M.~G. Kozlov and L.~N. Labzowsky.
\newblock Parity violation effects in diatomics.
\newblock {\em J. Phys. B}, 28(10):1933, 1995.

\bibitem{Nayak2006}
M.~K. Nayak and R.~K. Chaudhuri.
\newblock Ab initio calculation of {P, T} -odd interaction constant in {BaF}: a
  restricted active space configuration interaction approach.
\newblock {\em J. Phys. B}, 39(5):1231, 2006.

\bibitem{Fukuda2016}
M.~Fukuda, K.~Soga, M.~Senami, and A.~Tachibana.
\newblock Local spin dynamics with the electron electric dipole moment.
\newblock {\em Phys. Rev. A}, 93:012518, 2016.

\bibitem{Gaul2017}
K.~Gaul and R.~Berger.
\newblock Zeroth order regular approximation approach to electric dipole moment
  interactions of the electron.
\newblock {\em J. Chem. Phys.}, 147(1):014109, 2017.

\bibitem{Abe2018}
M.~Abe, V.~S. Prasannaa, and B.~P. Das.
\newblock Application of the finite-field coupled-cluster method to calculate
  molecular properties relevant to electron electric-dipole-moment searches.
\newblock {\em Phys. Rev. A}, 97:032515, 2018.

\bibitem{Haase2018}
P.~A.~B. Haase, M.~Ilia\v{s}, E.~Eliav, P.~Aggarwal, H.~L. Bethlem,
  A.~Borschevsky, M.~Denis, K.~Esajas, Y.~Hao, S.~Hoekstra, K.~Jungmann,
  T.~Meijknecht, M.~Mooij, R.~G.~E. Timmermans, W.~Ubachs, L.~Willmann, and
  A.~Zapara.
\newblock Provisional title: Electron edm: Relativistic coupled cluster
  calculations of {$W_d$} in {BaF}.
\newblock In preparation, 2018.

\bibitem{Osterwalder2010}
A.~Osterwalder, S.A. Meek, G.~Hammer, H.~Haak, and G.~Meijer.
\newblock {Deceleration of neutral molecules in macroscopic traveling traps}.
\newblock {\em Phys. Rev. A}, 81(5):051401, May 2010.

\bibitem{Bulleid2012}
N.~E. Bulleid, R.~J. Hendricks, E.~A. Hinds, S.~A. Meek, G.~Meijer,
  A.~Osterwalder, and M.~R. Tarbutt.
\newblock {Traveling-wave deceleration of heavy polar molecules in
  low-field-seeking states}.
\newblock {\em Phys. Rev. A}, 86(2):021404, 2012.

\bibitem{Quintero-perez2014}
M.~Quintero-P\'{e}rez, T.~E. Wall, S.~Hoekstra, and H.~L. Bethlem.
\newblock Preparation of an ultra-cold sample of ammonia molecules for
  precision measurements.
\newblock {\em J. Mol. Spectr.}, 300:112 -- 115, 2014.

\bibitem{Vandenberg2014}
J.~E. van~den Berg, S.~C. Mathavan, C.~Meinema, J.~Nauta, T.~H. Nijbroek,
  K.~Jungmann, H.~L. Bethlem, and S.~Hoekstra.
\newblock Traveling-wave deceleration of {SrF} molecules.
\newblock {\em J. Mol. Spectros.}, 300:22 -- 25, 2014.

\bibitem{Shuman2010}
E.~S. Shuman, J.~F. Barry, and D.~DeMille.
\newblock Laser cooling of a diatomic molecule.
\newblock {\em Nature}, 467:820--823, 2010.

\bibitem{Patterson2007}
D.~Patterson and J.~M. Doyle.
\newblock Bright, guided molecular beam with hydrodynamic enhancement.
\newblock {\em J. Chem. Phys.}, 126(15):154307, 2007.

\bibitem{Lim2017}
J.~Lim, J.~R. Almond, M.~A. Trigatzis, J.~A. Devlin, N.~J. Fitch, B.~E. Sauer,
  M.~R. Tarbutt, and E.~A. Hinds.
\newblock Laser cooled {YbF} molecules for measuring the electron's electric
  dipole moment.
\newblock {\em Phys. Rev. Lett.}, 120:123201, 2018.

\bibitem{Baron2017}
J.~Baron, W.~C. Campbell, D.~DeMille, J.~M. Doyle, G.~Gabrielse, Y.~V.
  Gurevich, P.~W. Hess, N.~R. Hutzler, E.~Kirilov, I.~Kozyryev, B.~R.
  O’Leary, C.~D. Panda, M.~F. Parsons, B.~Spaun, A.~C. Vutha, A.~D. West,
  E.~P. West, and ACME Collaboration.
\newblock Methods, analysis, and the treatment of systematic errors for the
  electron electric dipole moment search in thorium monoxide.
\newblock {\em New J. Phys.}, 19(7):073029, 2017.

\bibitem{Kara2012}
D.~M. Kara, I.~J. Smallman, J.~J. Hudson, B.~E. Sauer, M.~R. Tarbutt, and E.~A.
  Hinds.
\newblock Measurement of the electron's electric dipole moment using {YbF}
  molecules: methods and data analysis.
\newblock {\em New J. Phys.}, 14(10):103051, 2012.

\bibitem{Hudson2014}
J.~J. Hudson, M.~R. Tarbutt, B.~E. Sauer, and E.~A. Hinds.
\newblock {Stochastic multi-channel lock-in detection}.
\newblock {\em New J. Phys.}, 16, 2014.

\bibitem{Mathavan2016}
S.~C. Mathavan, A.~Zapara, Q.~Esajas, and S.~Hoekstra.
\newblock Deceleration of a supersonic beam of {SrF} molecules to 120 m/s.
\newblock {\em ChemPhysChem}, 17(22):3709--3713, 2016.

\bibitem{Hutzler2011}
N.~R. Hutzler, M.~F. Parsons, Y.~V. Gurevich, P.~W. Hess, E.~Petrik, B.~Spaun,
  A.~C. Vutha, D.~DeMille, G.~Gabrielse, and J.~M. Doyle.
\newblock A cryogenic beam of refractory{,} chemically reactive molecules with
  expansion cooling.
\newblock {\em Phys. Chem. Chem. Phys.}, 13:18976--18985, 2011.

\bibitem{Bulleid2013}
N.~E. Bulleid, S.~M. Skoff, R.~J. Hendricks, B.~E. Sauer, E.~A. Hinds, and
  M.~R. Tarbutt.
\newblock Characterization of a cryogenic beam source for atoms and molecules.
\newblock {\em Phys. Chem. Chem. Phys.}, 15:12299--12307, 2013.

\bibitem{Maxwell2005}
S.~E. Maxwell, N.~Brahms, R.~deCarvalho, D.~R. Glenn, J.~S. Helton, S.~V.
  Nguyen, D.~Patterson, J.~Petricka, D.~DeMille, and J.~M. Doyle.
\newblock {High-Flux Beam Source for Cold, Slow Atoms or Molecules}.
\newblock {\em Phys. Rev. Lett.}, 95(1):173201, 2005.

\bibitem{vanBuuren2009}
L.~D. van Buuren, C.~Sommer, M.~Motsch, S.~Pohle, M.~Schenk, J.~Bayerl,
  P.~W.~H. Pinkse, and G.~Rempe.
\newblock Electrostatic extraction of cold molecules from a cryogenic
  reservoir.
\newblock {\em Phys. Rev. Lett.}, 102(3):033001, 2009.

\bibitem{Skoff2011}
S.~M. Skoff, R.~J. Hendricks, C.~D.~J. Sinclair, J.~J. Hudson, D.~M. Segal,
  B.~E. Sauer, E.~A. Hinds, and M.~R. Tarbutt.
\newblock Diffusion{,} thermalization{,} and optical pumping of {YbF} molecules
  in a cold buffer-gas cell.
\newblock {\em Phys. Rev. A}, 83:023418, 2011.

\bibitem{Hutzler2012}
N.~R. Hutzler, H.-I. Lu, and J.~M. Doyle.
\newblock The buffer gas beam: {A}n intense, cold, and slow source for atoms
  and molecules.
\newblock {\em Chem. Rev.}, 112(9):4803--4827, 2012.

\bibitem{Zhou2015}
Y.~Zhou, D.~D. Grimes, T.~J. Barnum, D.~Patterson, S.~L. Coy, E.~Klein, J.~S.
  Muenter, and R.~W. Field.
\newblock Direct detection of {R}ydberg{$–$}{R}ydberg millimeter-wave
  transitions in a buffer gas cooled molecular beam.
\newblock {\em Chem. Phys. Lett.}, 640:124--136, 2015.

\bibitem{Cheng2016}
C.~Cheng, A.~P.~P. van~der Poel, P.~Jansen, M.~Quintero-P\'erez, T.~E. Wall,
  W.~Ubachs, and H.~L. Bethlem.
\newblock Molecular fountain.
\newblock {\em Phys. Rev. Lett.}, 117:253201, 2016.

\bibitem{Zieger2010}
P.~C. Zieger, S.~Y.~T. van~de Meerakker, C.~E. Heiner, H.~L. Bethlem, A.~J.~A.
  van Roij, and G.~Meijer.
\newblock Multiple packets of neutral molecules revolving for over a mile.
\newblock {\em Phys. Rev. Lett.}, 105:173001, 2010.

\bibitem{Fabrikant2014}
M.~I. Fabrikant, T.~Li, N.~J. Fitch, N.~Farrow, J.~D. Weinstein, and H.~J.
  Lewandowski.
\newblock Method for traveling-wave deceleration of buffer-gas beams of {CH}.
\newblock {\em Phys. Rev. A}, 90:033418, 2014.

\bibitem{Meerakker2012}
S.~Y.~T. van~de Meerakker, H.~L. Bethlem, N.~Vanhaecke, and G.~Meijer.
\newblock {Manipulation and control of molecular beams}.
\newblock {\em Chem. Rev.}, 112(9):4828--4878, 2012.

\bibitem{VandenBerg2012}
J.~E. van~den Berg, S.~H. Turkesteen, E.~B. Prinsen, and S.~Hoekstra.
\newblock Deceleration and trapping of heavy diatomic molecules using a
  ring-decelerator.
\newblock {\em Eur. Phys. J. D}, 66(9):235, 2012.

\bibitem{Quintero-perez2013}
M.~Quintero-P\'erez, P.~Jansen, T.~E. Wall, J.~E. van~den Berg, S.~Hoekstra,
  and H.~L. Bethlem.
\newblock Static trapping of polar molecules in a traveling wave decelerator.
\newblock {\em Phys. Rev. Lett.}, 110:133003, 2013.

\bibitem{Jansen2013}
P.~Jansen, M.~Quintero-P\'erez, T.~E. Wall, J.~E. van~den Berg, S.~Hoekstra,
  and H.~L. Bethlem.
\newblock Deceleration and trapping of ammonia molecules in a traveling-wave
  decelerator.
\newblock {\em Phys. Rev. A}, 88:043424, 2013.

\bibitem{Hummon2013}
M.~T. Hummon, M.~Yeo, B.~K. Stuhl, A.~L. Collopy, Y.~Xia, and J.~Ye.
\newblock {2D} magneto-optical trapping of diatomic molecules.
\newblock {\em Phys. Rev. Lett.}, 110:143001, 2013.

\bibitem{Zhelyazkova2014}
V.~Zhelyazkova, A.~Cournol, T.~E. Wall, A.~Matsushima, J.~J. Hudson, E.~A.
  Hinds, M.~R. Tarbutt, and B.~E. Sauer.
\newblock Laser cooling and slowing of {CaF} molecules.
\newblock {\em Phys. Rev. A}, 89:053416, 2014.

\bibitem{Barry2014}
J.~F. Barry, D.~J. McCarron, E.~B. Norrgard, M.~H. Steinecker, and D.~DeMille.
\newblock {Magneto-optical trapping of a diatomic molecule}.
\newblock {\em Nature}, 512(7514):286--289, 2014.

\bibitem{Kozyryev2017}
I.~Kozyryev, L.~Baum, K.~Matsuda, B.~L. Augenbraun, L.~Anderegg, A.~P. Sedlack,
  and J.~M Doyle.
\newblock {Sisyphus Laser Cooling of a Polyatomic Molecule}.
\newblock {\em Phys. Rev. Lett.}, 118(1):173201, 2017.

\bibitem{De2009}
S.~De, U.~Dammalapati, K.~Jungmann, and L.~Willmann.
\newblock Magneto-optical trapping of barium.
\newblock {\em Phys. Rev. A}, 79:041402, 2009.

\bibitem{Berg1993}
L.-E. Berg, T.~Olsson, J.-C. Chanteloup, A.~Hishikawa, and P.~Royen.
\newblock Lifetime measurements of excited molecular states using a
  {T}i:sapphire laser.
\newblock {\em Mol. Phys.}, 79(4):721--725, 1993.

\bibitem{Berg1998}
L.-E. Berg, N.~Gador, D.~Husain, H.~Ludwigs, and P.~Royen.
\newblock Lifetime measurements of the {A$^2\Pi_{1/2}$} state of {BaF} using
  laser spectroscopy.
\newblock {\em Chem. Phys. Lett.}, 287(1):89--93, 1998.

\bibitem{Effantin1987}
C.~Effantin, J.~D'Incan, A.~Bernard, G.~Fabre, R.~Stringat, J.~Verg{\`e}s, and
  R.~Barrow.
\newblock {Laser-induced fluorescence from {C$^2\Pi$} to {X$^2\Sigma^+$} and
  {H'$^2\Delta$} states of {BaF} analyzed by {F}ourier transform spectroscopy}.
\newblock {\em J. de Physique Colloques}, 48(C7):673--675, 1987.

\bibitem{Effantin1990}
C.~Effantin, A.~Bernard, J.~d'Incan, G.~Wannous, J.~Verg\`{e}s, and R.~F.
  Barrow.
\newblock Studies of the electronic states of the {BaF} molecule.
\newblock {\em Mol. Phys.}, 70(5):735--745, 1990.

\bibitem{Bernard1990}
A.~Bernard, C.~Effantin, J.~d'Incan, J.~Verg\`{e}s, and R.F. Barrow.
\newblock Studies of the electronic states of the {BaF} molecule.
\newblock {\em Mol. Phys.}, 70(5):747--755, 1990.

\bibitem{Bernard1992}
A.~Bernard, C.~Effantin, E.~Andrianavalona, J.~Verg\`{e}s, and R.~F. Barrow.
\newblock Laser-induced fluorescence of {BaF}: {F}urther results for six
  electronic states.
\newblock {\em J. Mol. Spectr.}, 152(1):174 -- 178, 1992.

\bibitem{Chen2016}
T.~Chen, W.~Bu, and B.~Yan.
\newblock Structure, branching ratios, and a laser-cooling scheme for the
  $^{138}\mathrm{BaF}$ molecule.
\newblock {\em Phys. Rev. A}, 94:063415, 2016.

\bibitem{Chen2017}
T.~Chen, W.~Bu, and B.~Yan.
\newblock Radiative deflection of a {BaF} molecular beam via optical cycling.
\newblock {\em Phys. Rev. A}, 96:053401, 2017.

\bibitem{Hao2018}
Y.~Hao, L.~F. Pa\v{s}teka, L.~Visscher, P.~Aggarwal, H.~L. Bethlem,
  A.~Borschevsky, K.~Esajas, P.~A.~B. Haase, S.~Hoekstra, K.~Jungmann,
  T.~Meijknecht, M.~Mooij, R.~G.~E. Timmermans, W.~Ubachs, L.~Willmann, and
  A.~Zapara.
\newblock Provisional title: Relativistic {F}ock-space coupled cluster and
  configuration interaction studies of spectroscopic properties of selected
  alkaline earth metal fluorides.
\newblock In preparation, 2018.

\bibitem{Barrow1988}
R.F. Barrow, A.~Bernard, C.~Effantin, J.~D'Incan, G.~Fabre, A.~{El Hachimi},
  R.~Stringat, and J.~Verg\`{e}s.
\newblock The metastable {A$^{\prime 2}\Delta$} state of {BaF}.
\newblock {\em Chem. Phys. Lett.}, 147(6):535--537, 1988.

\bibitem{Kang2016}
S.~Kang, F.~Kuang, G.~Jiang, and J.~Du.
\newblock The suitability of barium monofluoride for laser cooling from ab
  initio study.
\newblock {\em Mol. Phys.}, 114(6):810--818, 2016.

\bibitem{Ernst1986}
W.~E. Ernst, J.~K\"{a}ndler, and T.~T\"{o}rring.
\newblock Hyperfine structure and electric dipole moment of {B}a{F}
  {X$^2\Sigma^+$}.
\newblock {\em J. Chem. Phys.}, 84(9):4769--4773, 1986.

\bibitem{Paperno2000}
E.~Paperno, H.~Koide, and I.~Sasada.
\newblock A new estimation of the axial shielding factors for multishell
  cylindrical shields.
\newblock {\em J. Appl. Phys.}, 87(9):5959--5961, 2000.

\bibitem{Budker2007}
D.~Budker and M.~Romalis.
\newblock Optical magnetometry.
\newblock {\em Nat. Phys.}, 3:227--234, 2007.

\bibitem{Allmendinger2017}
F.~Allmendinger, P.~Bl{\"u}mler, M.~Doll, O.~Grasdijk, W.~Heil, K.~Jungmann,
  S.~Karpuk, H.-J. Krause, A.~Offenh{\"a}usser, M.~Repetto, U.~Schmidt,
  Y.~Sobolev, K.~Tullney, L.~Willmann, and S.~Zimmer.
\newblock Precise measurement of magnetic field gradients from free spin
  precession signals of $^3${H}e and $^{129}${Xe} magnetometers.
\newblock {\em Eur. Phys. J. D}, 71(4):98, 2017.

\bibitem{Grasdijk2018}
O.~Grasdijk.
\newblock Phd thesis, {University of Groningen}.
\newblock In preparation, 2018.

\end{thebibliography}
\end{document}